\documentclass[]{llncs}

\usepackage{hyperref,cleveref,enumitem,url,tabularx}
\usepackage[misc]{ifsym}
\usepackage{graphicx} 
\newcommand{\CE}{Continuous Experimentation}
\newcommand{\cps}{cyber-physical systems}

\title{Continuous Experimentation for \\Automotive Software on the Example of a \\Heavy Commercial Vehicle in Daily Operation} 

\author{Federico Giaimo\inst{1}\orcidID{0000-0003-2400-1465}\and
Christian Berger\inst{2}}
\institute{Chalmers University of Technology, Gothenburg, Sweden\\
\email{giaimo@chalmers.se}\and
University of Gothenburg, Gothenburg, Sweden\\
\email{christian.berger@gu.se}}

\usepackage{tikz}

\newcommand\copyrighttext{%
  \footnotesize \textcopyright~Accepted for publication by the 14th European Conference on Software Architecture (ECSA 2020)}
\newcommand\copyrightnotice{%
\begin{tikzpicture}[remember picture,overlay]
\node[anchor=south,yshift=30pt] at (current page.south) {\fbox{\parbox{\dimexpr\textwidth-\fboxsep-\fboxrule\relax}{\copyrighttext}}};
\end{tikzpicture}%
}

\begin{document}

\maketitle
\copyrightnotice

\begin{abstract}
As the automotive industry focuses its attention more and more towards the software functionality of vehicles, techniques to deliver new software value at a fast pace are needed. 
Continuous Experimentation, a practice coming from the web-based systems world, is one of such techniques.
It enables researchers and developers to use real-world data to verify their hypothesis and steer the software evolution based on performances and user preferences, reducing the reliance on simulations and guesswork. 
Several challenges prevent the verbatim adoption of this practice on automotive cyber-physical systems, e.g., safety concerns and limitations from computational resources; nonetheless, the automotive field is starting to take interest in this technique. 
This work aims at demonstrating and evaluating a prototypical Continuous Experimentation infrastructure, implemented on a distributed computational system housed in a commercial truck tractor that is used in daily operations by a logistic company on public roads. 
The system comprises computing units and sensors, and software deployment and data retrieval are only possible remotely via a mobile data connection due to the commercial interests of the logistics company. 
This study shows that the proposed experimentation process resulted in the development team being able to base software development choices on the real-world data collected during the experimental procedure. 
Additionally, a set of previously identified design criteria to enable Continuous Experimentation on automotive systems was discussed and their validity confirmed in the light of the presented work. 


\end{abstract}

\section{Introduction}
The automotive industry is currently investing considerable efforts and resources towards the achievement of an autonomous vehicle that would meet the specification of SAE level 3~\cite{sae}. 
Several companies have in fact already marketed vehicles exhibiting different semi-autonomous capabilities belonging to SAE level 2, ranging from adaptive lane keeping to self-parking features. 
The most relevant difference between level 2 and 3 in the SAE hierarchy is on who takes the responsibility of monitoring the driving environment: while in SAE level 2 the system assists the human driver in latitudinal and longitudinal adjustments, it is the driver who is expected to perform all the remaining tasks; instead, in SAE level 3 this is not required, meaning that the vehicle itself should be able to manage the \emph{dynamic driving tasks} while the human driver is only expected to intervene upon request~\cite{SAE_levels}. 

The software necessary to manage the diversity of situations that a vehicle can face is bound to be complex and computationally intensive, especially considering that the software present in modern vehicles already exceeds the Gigabyte in size~\cite{FUSE_Hiller}. 
Moreover, as all vehicles share the same basic capabilities but differ in the provided software functionality, it can be expected that the latter will constitute for the customers a relevant practical difference between automakers, thus fuelling a \emph{functionality race} that will move the value concentration from the hardware, i.e.,~the vehicle itself, to its software capabilities. 

\subsection{\CE}
When the software takes over the competitively distinguishing role from the hardware in the value-creation process, delivering new updates and functionality in a quick manner becomes necessary. 
This is very apparent in the software industry, especially for what concerns web-based software, where development techniques have been introduced to accelerate the process as much as possible by learning from how users and customers interact with such systems.
Among them we find Continuous Integration (CI), Continuous Deployment (CD), and \CE\ (CE). 

Continuous Integration proposes the integration of new software into the rest of the code base as soon as possible while Continuous Deployment involves the possibility of immediate deploying newly integrated software code into the target systems when all automated testing is successfully completed. 
There are many platforms that enable these two methodologies for software development teams, e.g., GitLab, Jenkins, and Zuul among many others. 
\CE\ builds upon the CI/CD pipeline and aims at enabling the developers to test new software performances by providing 
the possibility of deploying and running alongside the \emph{official} software a number of \emph{experiments}. 
These experiments could be either different versions of the official software or new functionality to be field-tested. 
While it adds computational overhead to the systems, \CE\ allows to confirm or reject hypotheses about the software suitability for a given task based on real-world data as opposed to simulations or speculations, making the software evolution process \textit{data-driven}. 

\CE\ has proven to be very effective on web-based software systems~\cite{GKT+19}. 
However, applying verbatim this way of working onto safety-critical cyber-physical systems such as vehicles would be an endeavour destined to face the specific challenges of the automotive context. 
One challenge is the added complexity given by the fact that the target systems in the case of vehicles are not virtual machines in server farms but highly mobile physical objects with limited computational performance. 
Moreover, there is a resource availability problem given by the \CE\ practice itself, which introduces a non-traditional approach when it comes to testing new functionality and needs additional computational power in order to manage the additional experiments and the data collection~\cite{GBK17} on top of the system's nominal functions. 
This can pose issues to the automotive industry, which, being based on an economy of scale, has always built vehicles with hardware that is \emph{just enough} powerful to fulfill its tasks in order to lower production costs. 
It also requires a rethinking of the classic system and software's architectures due to the new practice in which extra software is downloaded and run while its results are collected and uploaded back to the manufacturer. 
Nonetheless, new competitors seem to embrace this challenge as it can be seen from a manufacturer for luxury electric vehicles. 
In their quarterly financial reports, they mention already since 2015 the systematic gathering of driving and sensor data via ``field data feedback loops'' that are used to ``enable the system to continually learn and improve its performance''~\cite{tesla_quarterly}. 
While software experiments are not explicitly named, a company representative did mention the practice of installing ``an `inert' feature on vehicles'' in order to ``watch over tens of millions of miles how a feature performs'' by logging its behavior in a real-world scenario~\cite{tesla_inert}. 

A previous investigation in the automotive field by the authors shows that practitioners expect that they would benefit from the introduction of the \CE\ practice, even if it now faces these additional challenges~\cite{GAB19}. 
Another recent study showed that literature was generally focusing increasing efforts in the study of this practice, but only a small portion of these studies were actually proposing practical experiments and none of them in the context of a \CE\ setting on an automotive or \cps\ where the object of the experiment was not a visual change in a user interface~\cite{RR18}. 
Hence, the current work was devised to fill this research gap being the first study of this kind to propose and evaluate a system based on a proof-of-concept architecture for \CE\ built on previously identified design criteria~\cite{GB17}, housed on a commercial truck tractor operated on a daily basis by a logistics company in Sweden (the truck is still in use throughout 2020). 

\subsection{Scope of this work}
While the aim of this work is to draw conclusions that are valid for the automotive field, it is worth mentioning the differences between the experimental work and a commercial automotive scenario. 
One such scenario would generally involve a fleet of vehicles, likely passengers cars, which are each controlled by a number of highly resource-constrained Electronic Control Units (ECUs). 
The experimental work for this study was instead performed on a single vehicle, i.e., a commercial truck tractor, equipped with a server-grade computing unit more powerful than a typical ECU, and the software was written using a high-level programming language. 
These differences are due to the fact that the aim of this study is \emph{to provide and evaluate a proof-of-concept for the \CE\ process} rather than focusing on a particular automotive function. 
A key aspect is however preserved: in the real-world case and in this study the vehicle is physically inaccessible to the manufacturer, forcing all software deployment and data exchange to be performed via an Over-The-Air (OTA) connection while the vehicle is in operation. 
Finally, it should be noted that the scope of this study does not include autonomous driving tasks as the vehicle used in the experimental setting is manually driven by a driver from the logistics company. 

\subsection{Research Goal}\label{sec:rg}
Previous investigations clearly show that the literature lacks design science studies about \CE\ in realistic cyber-physical systems contexts, and especially in the automotive domain. 
This study aims to bridge this research gap. 
The Research Goal (RG) of this work can be expressed as: 
\begin{enumerate}[leftmargin=*,align=left,label=\emph{RG}:] 
\item To provide and evaluate a proof-of-concept that shows the feasibility and benefit of a \CE\ decision cycle for an algorithmic choice in the context of an automotive system, based on previously identified design criteria. 
\end{enumerate} 
The Research Goal of this article can be further divided in the following Research Questions (RQ):
\begin{enumerate}[leftmargin=*,align=left,label=\emph{RQ{\arabic*}:}] 
\item What software architecture can support a \CE\ decision process on a complex \cps\ such as an automotive system? 
\item To what extent do previously identified design criteria for \CE\ in the context of automotive \cps\ hold?
\end{enumerate}

\subsection{Contributions}
To the best knowledge of the authors, this study presents for the first time a \CE\ decision cycle focused on an algorithmic experiment on a computational system housed in a commercial vehicle, where the deployment of experimental software to the system and the retrieval of gathered data are performed via a mobile data connection while the automotive system was operated by the owner company. 
The whole experimental setting aimed to be the least invasive for the company's operators and their commercial activities. 
Both the system and software architectures are reported and the experimental work offered the chance to discuss and validate a set of design criteria for \CE\ on automotive \cps\ that were previously identified in a preceding study.

\section{Related Works}\label{sec:rw}

A number of studies explore the \CE\ practice, in its native application field, i.e., web-based systems, and more recently in the context of \cps. 
Gupta \emph{et al.}~\cite{GKT+19} describe the First Practical Online Controlled Experiments Summit. 
During this summit, a number of experts in experimentation from several software and online-based companies convened to discuss the experimentation processes they have in place, the main challenges they are facing, and some relative solutions. 

Fagerholm \emph{et al.}~\cite{FGMM17} defined an organizational model for \CE\ in the context of web-based products, comprising the tasks and artefacts that different roles involved in planning and implementation of a software product should manage in order to enable the experimentation process.

Recent mapping studies on the \CE\ practice show that the majority of the works they encountered explore the statistical methods sub-topic and are often rooted in the web-based applications context, which is the originating field of this practice; only a minority of studies are addressing the \CE\ practice in the \cps\ field~\cite{AF18,RR18}.

A previous work led by the authors~\cite{GB17} explored the design characteristics that a \cps\ should possess in order to enable a \CE\ process on an autonomous vehicle. 
These design criteria are evaluated in this study to discuss their validity in the light of the presented work and considering the difference between the scopes of the two studies. 

Olsson and Bosch~\cite{OB13} published a study connecting post-deployment data and the cyber-physical and automotive field. 
They interviewed representatives from three companies, one of which is an automotive manufacturer. 
The study reports that while post-deployment data collection mechanisms are in place, the collected data is only partially used and the purpose of this feedback is troubleshooting, rather than supporting a product improvement process. 

Mattos \emph{et al.}~\cite{MBO18} performed a literature review to identify a set of challenges for \CE\ in \cps\, that was used a starting point for a case study where they tried to identify possible solutions with industrial representatives. 

Cioroaica \emph{et al.}~\cite{CKB19} propose the analysis of Digital Twins to assess the trustworthiness of smart agents such as additional functionality or system component being downloaded to a smart vehicle. 
While the approach yields value especially to evaluate third-party functionality, it relies on simulating the new component's behavior in a partial simulation of the surrounding environment. 
While simulations should be part of the evaluation process for new software due to the safety they can guarantee, in the authors' view they cannot completely replace the value coming from a field evaluation since the very high complexity of the real world and the system's interaction with it cannot be perfectly simulated. 

No relevant publicly available information was found about commercial companies' practices regarding internal software experiments to improve autonomous functionality, except from the aforementioned comments regarding inert features~\cite{tesla_quarterly,tesla_inert}. 

\section{Methodology}\label{sec:meth}

A Design Science methodology, i.e.,~the design and investigation of artifacts in context~\cite{W14}, was adopted to achieve the Research Goal. 
A software architecture was devised to support a number of software modules that would run and interact on a system performing a \CE\ decision cycle, housed in a commercial heavy vehicle, shown in~\Cref{fig:voyager}. 
The \CE\ practice was applied to answer in a data-driven fashion a software development question regarding an algorithmic choice, performed on a complex \cps\ such as an automotive vehicle only accessible via a remote connection. 
While supporting a software experiment is the goal, the focus of this study is not on the experiment itself, i.e., what the production and experimental modules actually do, but instead on the experimentation process itself. 
In other words, even if an experiment has been set up, for the purpose of this study what matters is not the result of the experiment, but rather whether an experiment could be actually carried out according to the \CE\ practice. 
For this reason, the focus of the results and discussion is the architecture and infrastructure for the experiment and not its outcome. 

The experiment consisted in running different Machine Learning-based object detectors connected to the live video feed in order to find an object detector module that would recognise, as accurately as possible, items and road users in the vehicle's field of view. 
The experiment was run in a series of time-wise short sessions and the resulting data were analysed manually. 
The machine learning software modules were based on publicly available detection models\footnote{\url{https://github.com/tensorflow/models/blob/master/research/object\_detection/g3doc/detection_model_zoo.md}} pre-trained on the COCO dataset~\cite{coco}. 
This dataset was chosen because of the breadth of its scope, which encompasses automotive items and more, making it a valuable choice for a general-purpose object detector.

\section{Results}\label{sec:res}

\subsection{Research Question 1}
The work here reported shows a system and software architecture for the application of a \CE\ methodology in order to answer a software development question regarding an algorithmic choice, on a system housed in a remotely accessible vehicle. 
The following paragraphs describe the details of the software architecture supporting the experimentation process, the system architecture enabling the software to gather data and communicate results, and the way that the software was packaged in order to ease the deployment process while following the Continuous Integration/Continuous Deployment practices. 

\subsubsection{Software Architecture}
\begin{figure}[thpb]
\centering
\includegraphics[width=0.85\linewidth]{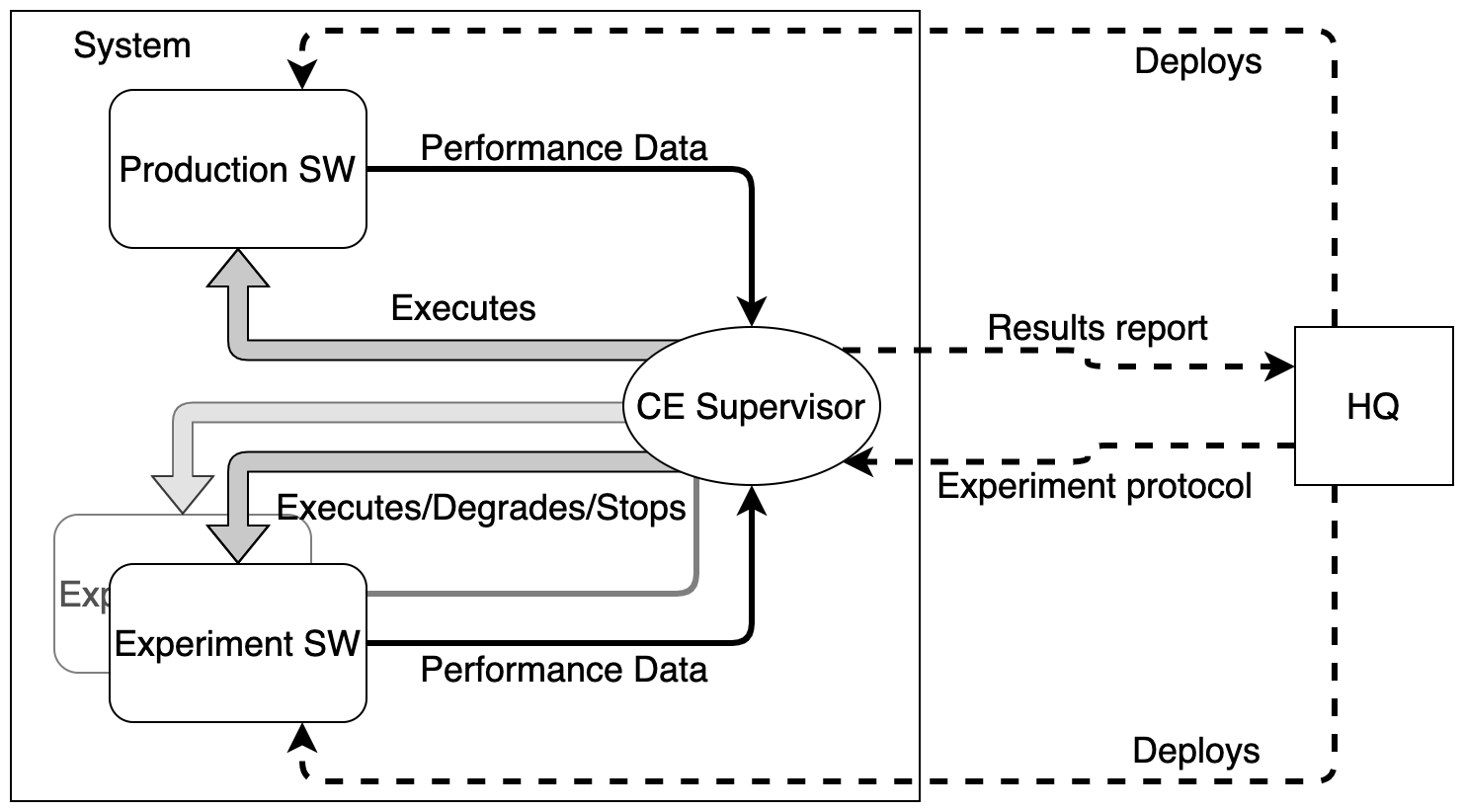}
\caption{View of the system and its components. The dashed lines represent Over-The-Air (OTA) communication.}
\label{fig:CE}
\end{figure}

The experimentation process is based on the interaction of the three modules \emph{Production Software}, \emph{Experimental Software}, and \emph{Supervisor}, as shown in~\Cref{fig:CE}. 
As the names suggests, Production Software simulates a production component, whose performance must not be influenced by any other components. 
Each instance of the Experimental Software module represents an experiment deployed to test a new software variant, which runs in a sandboxed way, i.e.,~they must not issue commands to the actual system (especially any actuators) but instead have their output logged for later analysis, similarly to what is done by an automotive manufacturer who revealed it uses ``inert features''~\cite{tesla_inert}. 
The Supervisor module poses as the experiment manager software, monitoring the other modules' performances and deciding at any time whether to continue or not with the experiment, depending on whether the Experimental Software modules abide to the experiment parameters. 
It is also the module that interacts with the team, represented by the ``HQ'' box in~\Cref{fig:CE}, that plans and conducts the deployment of both the software modules and the Experiment Protocol, which comprises the parameters of the experiment cycle. 
Finally, it reports the results observed during operation back to the team. 

When an experiment is set up in the computing system, an \emph{Experimentation Protocol} is provided, which is a file collecting relevant parameters for the experiment, e.g., CPU usage thresholds for the Experimental Software that should not be crossed. 
Upon starting, the Supervisor will wait for the other software modules to manage the experimentation process. 
If a performance drop in the Production Software or an increase in resources consumption by the Experimental Software modules is detected by the Supervisor, the change is compared to the thresholds as specified in the Experimentation Protocol. 
If necessary, the Supervisor has the capability to request the Experimental Software modules either a \emph{performance degradation}, so that it consumes less resources thus leaving more for the Production Software, or a full stop of the experiment if the violations are deemed too severe. 
During the experiment, relevant data about the detection performances are collected. 
These results are transmitted back to the remote team at the end of each experiment, allowing them to analyze the experiment's performance and finally decide which software version fulfilled its functional objectives more effectively.

\subsubsection{System Architecture}\label{sec:arch}
To provide a proof-of-concept for Continuous Experimentation in the automotive context and better understand the underlying challenges, a research project was initiated as a collaboration between Chalmers University of Technology's vehicular laboratory REVERE, Volvo, Trafikverket, GDL, Kerry Logistics, Speed Group, Borås Stad, Ellos, and Combitech to equip a modern Volvo tractor with a platform consisting of two computers, five cameras, three GPS sensors, and a GPS/IMU system for daily data logging during typical operations of a logistics company. 

\begin{figure}[thpb]
\centering
\includegraphics[width=0.85\linewidth]{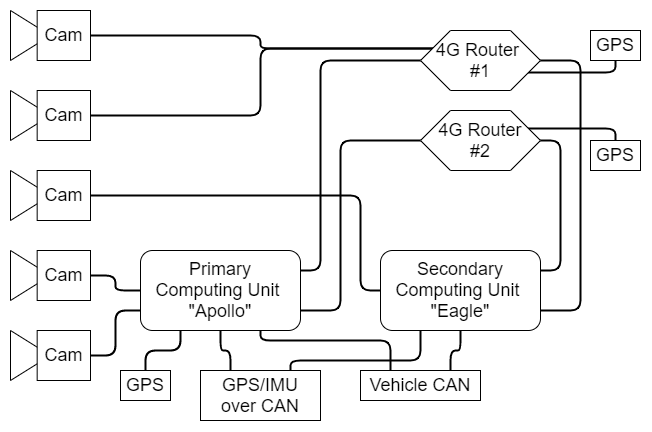}
\caption{Architecture of the system, named \emph{Voyager}. 
The two computing units provide some redundancy should a fail occur while the system is in mission.
A comprehensive set of cameras is available, as well as an IMU signal and several GPS sources.}
\label{fig:AF}
\end{figure}

As depicted in~\Cref{fig:AF}, the system is designed in the following manner: 
The automotive platform, a commercial truck tractor, is equipped with a Linux-based, Docker-enabled computer as primary unit and an Accelerated Processing Unit (APU) as secondary computing node. 
The main computer is equipped with an Intel Core i9-9900K CPU and an NVidia GP107 GPU. 
It is directly connected to two cameras, two GPS systems, and the vehicle's CAN network. 
The secondary unit has instead direct access to one camera, one GPS unit, and the vehicle's CAN network, since the computing systems are capable to access a subset of the CAN signals of the automotive platform, specifically the ones containing the vehicle's speed and the IMU data.
The secondary unit has the purpose of providing a stable, low-energy demanding, highly available connection, enabling an additional point of access to the system for maintenance purposes. 
Moreover, being directly connected to a number of input sensors, it can also act as a reliable fail-over system, although with degraded performances and a reduced amount of data, should the main unit malfunction during operations. 
Finally, the two mobile data connection routers acting as internal network nodes connect internally both computing units to the remaining two cameras and GPS units, and externally the whole system to the outside world.
To provide a stable power supply to the hardware and not limit operations to only the time when the engine runs, the system is powered by a battery pack which is recharged by the engine when it is running. 

\begin{figure}[thpb]
  \centering
  \begin{minipage}[b]{0.465\textwidth}
    \includegraphics[width=\textwidth]{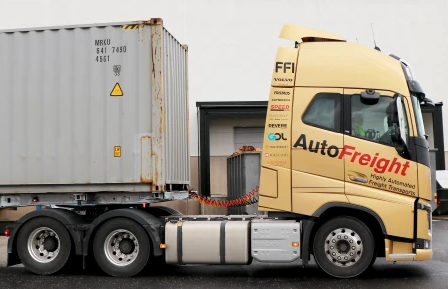}
    \caption{The vehicle housing the system is part of a project named Highly Automated Freight Transports (AutoFreight).}
    \label{fig:voyager}
  \end{minipage}
  \hfill
  \begin{minipage}[b]{0.465\textwidth}
    \includegraphics[width=\textwidth]{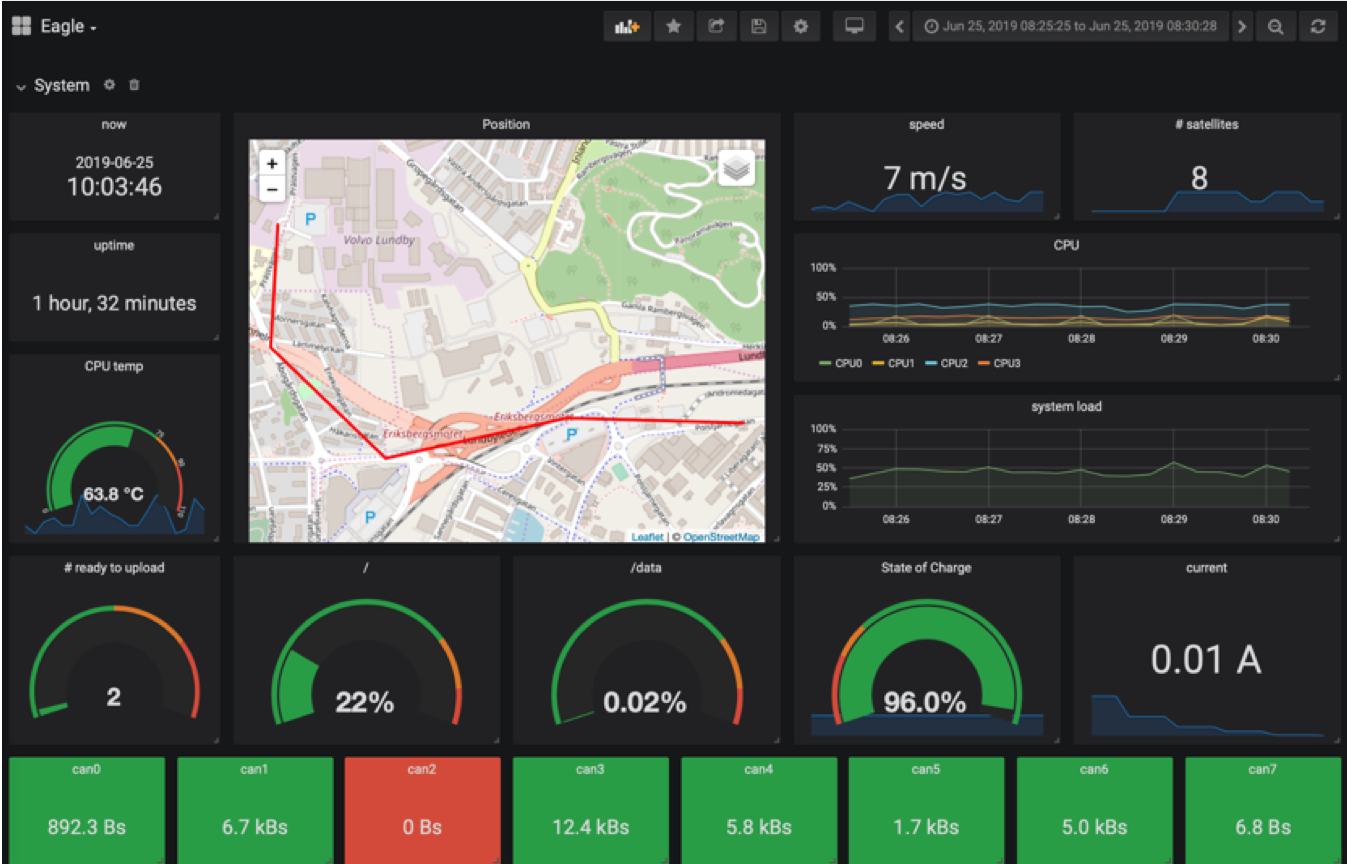}
    \caption{Dashboard monitoring one of the computing systems on board the commercial truck tractor used in the experimental phase of this work.}
    \label{fig:dashboard}
  \end{minipage}
\end{figure}

The system is monitored live through a software dashboard, shown in~\Cref{fig:dashboard}, that allows to easily visualize important parameters such as system time, up-time, CPU temperature and consumption, system load, vehicle speed, GPS position and number of satellites, storage disk space utilization, battery level, and CAN connections data rates.
As the vehicle is in daily operation by the logistics company, the software to test and the resulting data can only be extracted remotely hence making this project and platform well-suited for this study on \CE, as it represents the use-case of a single vehicle in a fleet that can run software experiments but cannot be physically accessed by the manufacturer. 

\subsubsection{Software Development and Deployment}
To simplify the deployment phase, all software modules were developed and encapsulated using Docker\footnote{\url{https://www.docker.com}}. 
Docker uses OS-level resource isolation to enable the execution of software in environments called containers, which are run on top of the host OS kernel, thus resulting in a more lightweight solution than a full-stack virtualization software. 
Each container is an instance of a Docker image, which acts similarly as a guest machine template and can be used to store and deliver applications. 

The versioning, integration, and deployment operations were run in a GitLab-based environment. 
GitLab is a web-based DevOps lifecycle environment that provides a Git repository manager providing, among the other services, a Continuous Integration/Continuous Deployment pipeline. 
The resulting development cycle would follow these steps:
firstly, a new change is introduced in the codebase via a Git repository commit; 
then, the Continuous Integration pipeline is automatically triggered and the new code is integrated and built within the code base; 
finally, the Continuous Deployment pipeline is executed triggering the build a Docker image, which is ready for distribution. 
If the new code was part of a software experiment, at the end of these three steps a Docker image with the experimental software is ready to be deployed and executed.
These steps embody what we can expect an industrial \CE\ cycle to look like from development to deployment to execution and finally, by instrumenting the code, to data collection, analysis, and choice of a final software variant. 

\begin{figure}[thpb]
\centering
\includegraphics[width=0.85\linewidth]{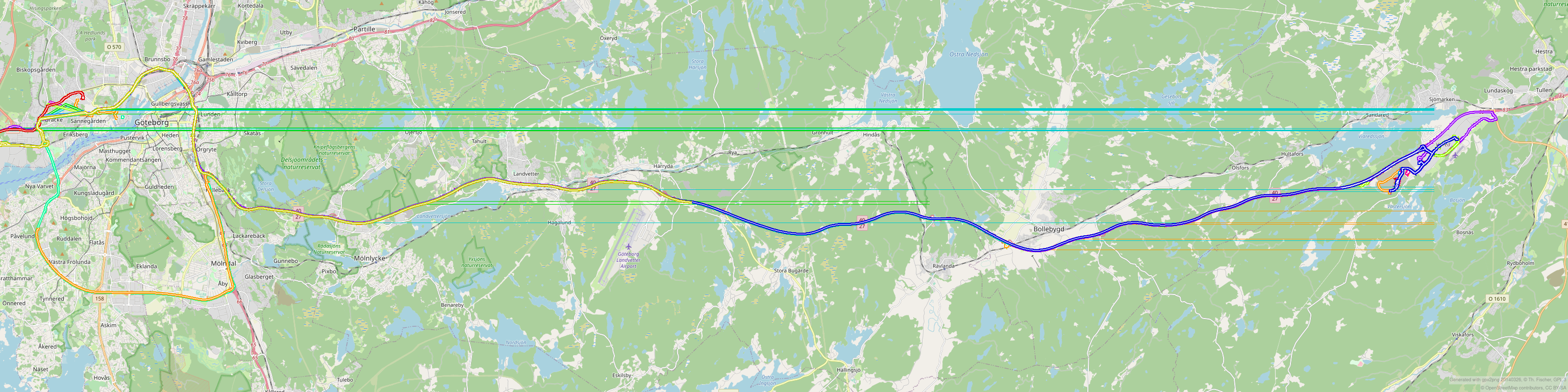}
\caption{Highlighted GPS traces of the vehicle in the first half of 2020 in the Gothenburg geographic area. The horizontal lines were artefacts of the overlay script in correspondence of GPS-denied areas.} 
\label{fig:gps}
\end{figure}

From what resulted during the development work, the average code base change would take around 4 minutes to be integrated while the Docker image building phase would last around 7 minutes. 
This means that a little more than ten minutes after new code was committed to the code base it was already available for deployment into the system. 
These phases took place at the team's end of the process and not on the automotive system itself, which had to download the software modules over the mobile connection. 
In the described setup, the resulting Docker image for an Experimental Software module amounted to approximately 5 GB in size due to the machine learning models and dependencies. 
While its size is significant, it is worth noting that no optimization nor compression was applied to the Docker image, which could have reduced significantly the amount of data to be deployed. 
The download of this image into the automotive system took approximately 14 minutes, which is comparable to the time needed to perform software updates in commercial vehicles. 
However, since Docker images are built as the aggregation of ID-marked layers based on their building process, most of the image downloads were only partial as several intermediate layers did not change between software builds and Docker allows to skip downloading duplicate layers. 
The experiment was run in a series of rounds while the vehicle was in operation in the Gothenburg area, as shown in~\Cref{fig:gps}. 
At the end of the experiment the resulting data were manually analyzed and it was concluded that the object detector used in one of the Experimental Software modules performed more accurately than the Production Software module. 
The results of the machine learning experiment are not reported nor discussed in detail as the focus of this work is not the object of the experiment in itself, but rather the architecture and infrastructure that made it possible. 
In the described experimental setup the process proved to be possible and feasible, and led to a successful experiment cycle that produced a data-based answer to a software development question.

\subsection{Research Question 2}
In a previous study on the subject a set of design properties was identified that would enable \CE\ on a complex cyber-physical system such as an autonomous vehicle~\cite{GB17}. 
These properties are here listed and discussed in the light of the work described so far.

\emph{Access to perception sensors and systems}, this was of course needed to run the Production and Experimental Software and was used in this study;
\emph{access to full vehicle control}, in this work it was not needed since controlling the vehicle was not in the scope of the experimentation process nor the experiment itself. Had it been so, a system architecture capable of driving the vehicle would have been needed; 
\emph{log internal activity and other relevant metrics}, a necessary step to allow the analysis of the experimental results;  
\emph{enabling of data transmission from the developers to the deployed system} and \emph{the feedback loop in the opposite direction}, also necessary to deploy software and retrieve the resulting data remotely;
\emph{reliability}, implemented through health checking techniques adopted to limit fault propagation and to enable remote troubleshooting and ``graceful degradation'' by having a secondary computing unit capable to restart the primary one and having access to own sensors and data streams; 
\emph{testability}, as all changes in functionality were firstly tested on local machines fed with recordings of past camera streams to ensure that the new code to be deployed to the system would perform as expected;
\emph{safety}, in this case the software had no physical control over the actual vehicle, meaning that even in case of faults, the safety implications were limited. 
Nonetheless, safety constraints were implemented in the form of thresholds over the amount of computational power that the experiment modules could use in order to simulate how the system would respond to resource-hungry experiments endangering the execution of Production Software; 
\emph{scalability}, an automotive system is naturally distributed across several computational units, in the present case the system adopted in this study is distributed over two computing nodes. 
While one was used to actually execute Production and Experimental Software modules, the other was still involved in the process as it was accessed to retrieve the camera feed used by the software.
Would it have been possible or necessary, the modular nature of the software that was used would have allowed for even more spread-out distribution, since the communication between software modules was performed via UDP multicast message exchange, requiring simply a network connection among computing nodes;
\emph{separation of concerns}, meaning the establishment of abstraction layers between hardware and software and between data and exchanged messages, definitely a necessary part of any software running on complex \cps; 
\emph{simplicity to involve new developers}, a feature of the development process more than of the physical system itself, in this case provided mostly by the ease of use of the development tools, which automated the majority of the steps necessary to perform Continuous Integration/Deployment pipelines;
\emph{facilitation for operators}, meaning that the software should not be hard to operate for those who are not developers, in this study it was not possible to acquire an external perspective on this point, as the only tester and operator of the \CE\ cycle was also the developer. However it should be noted that the adoption of microservices allowed to run or stop the execution of Production or Experimental Software by using a very limited number of console commands;
\emph{short cycle from development to deployment}, which is necessary whenever possible in order to roll out changes and new features at a fast pace, was definitely present in this study due to the automated Continuous Integration/Deployment mechanisms.

\section{Discussion}\label{sec:dis} 

The presented \CE\ prototypical implementation shows that it is possible to achieve enough data feedback from candidate functionality in a vehicular system to get a better understanding about its performances.
This allows researchers and developers to decide how to proceed with future software development efforts based on the data coming from the automotive system operating in real-life scenarios. 
As the goal was to verify the viability of the approach and qualitatively evaluate its architecture, the practical limits to the applicability or performances of the prototype, such as for example the minimum quality of service for the data connection or the base amount of experiments' results to be collected, were not in the direct focus of this study. 
Nonetheless it can be expected that certain parameters would be particularly relevant for the execution of the envisioned process, such as the remote connection quality, which has to be high enough to allow the exchange of software and the resulting data in the timeframe set for the experiment; and the computational capacity of the unit running the experiments, which has to support their execution so that the results of interest can be obtained. 

Since this was a proof-of-concept implementation, some of the issues that are specific to commercial vehicles were not addressed in this study. 
One of them is connected to the computational limits of automotive ECUs, which were not used in the experimental setup but are envisioned to be the computational units of such a production system in the future. 
Since ECUs are less computationally powerful than the hardware that was used in this prototype, employing them as computational hardware could have provided additional insight on how much could the low resources of these units hinder the execution of experiments. 
It is however worth to mention that even with low hardware capabilities it could be possible to run additional software, although perhaps not by using an off-the-shelf solution like Docker as it requires support from the Linux kernel. 
However, if adding additional computing power to the system is not an option it may still be possible to find scheduling strategies for the experiments' execution that make use of computational resources not needed by the Production Software~\cite{GBK17}. 

Another important difference between this prototype and commercial vehicles involves the safety constraints for the software. 
Automotive regulations demand strong safety standards for the software run in vehicles to which future experimental software may have to abide. 
In this prototype the only safety measures relied in the monitoring capabilities of the Supervisor module and its degradation/abort commands. 
Moreover, additional coding rules that apply to automotive software were not followed in this case, e.g., the prohibition to allocate dynamic memory. 
While sufficient for the aim of this test, it can be envisioned that more sophisticated coding standards and functional emergency stop mechanisms will be needed for future commercial implementations of this concept, unless perhaps it can be proved that the experiments cannot influence the vehicle's behavior in any way. 
Additional smaller challenges were posed by practical issues such as the size of software downloads to be undertaken by the automotive system, which was slowed by the bandwidth of the mobile data connection of the system. 
It should finally be mentioned that being this a prototype and not a system ready or close to commercial use, the company owner of the truck did not use the results of the study to change their strategy or operations at the present time.

Analyzing the design criteria identified in a previous study, it is the authors' conclusion that they do hold for a \CE\ process on an automotive system, with the only discrepancies explained by the lack of autonomous capabilities in the present study's vehicle and the presence of a single developer/tester instead of different team members covering different roles.
The design criteria can thus be viewed as a form of checklist to validate the preparedness of a complex \cps' architecture and development process to run \CE. 

\subsection{Threats to Validity}
A number of factors may threaten the validity of this work. 

One threat is likely the fact that the experiment infrastructure and the software modules do not abide to current automotive standards like~\cite{URL_iso21448,URL_iso26262}.
For example, one of the main differences between the software used in this work and the commercial automotive software is the use of dynamic memory allocation, which is currently forbidden in safety-critical systems due to the introduced vulnerability that could disrupt critical software capabilities when needed. 
This threatens the generalizability of the result since what was achieved in this study could be technically harder to obtain abiding to the strict automotive software standards. 
However, this threat is less impending considering that this work had the goal of providing a proof-of-concept showing that a working \CE-enabled vehicle is within the automotive industry's grasp, rather than provide one ready for commercial use.

Connected to the aforementioned threat, another potential issue is the fact that the software developed for this work had the capability to only run one or two Experimental Software modules at the same time. 
While this may seem an important limitation, it is worth noting that a higher number of experiments running concurrently would require a higher amount of spare computational power in a real-world scenario. 
Moreover, if a vehicle can only run a set amount of experiments at the same time this could play in favor of the development efforts necessary to tackle the previously mentioned threat to validity: the variables that would normally require an amount of memory dependent on the number of experiments could be in fact dimensioned \emph{a priori} since the number is fixed. 

Lastly, it should be noted that it is not necessarily possible to generalize the results obtained with \CE\ in the automotive field to the rest of the \cps\ context. 
While the challenges lurking in the automotive field are increasingly recognized and faced, it is possible and not unlikely that several additional challenges peculiar to different \cps\ sub-fields are still in the way and will prevent a rapid widespread adoption of this practice to non-automotive systems. 

\addtolength{\textheight}{-7.8cm}
\section{Conclusions and Future Work}\label{sec:cfw}
The presented work demonstrated and evaluated the execution of a prototypical \CE\ cycle for an automotive system, which is in daily commercial operations by a logistics company. 
The system was equipped with computing units and sensors and accessed remotely via a mobile connection, which was the only communication channel used to deploy software and retrieve the data resulting from running a software experiment. 
A set of previously identified design criteria to enable \CE\ on autonomous vehicles was discussed in light of the (non-autonomous) system built for this work. 
This study could show for the first time that an algorithmic development question can be answered applying a \CE\ process, while also highlighting some relevant challenges still standing on the way towards a fully-functional experiment-enabled vehicle. 

One direction for future studies could be for example the automation of those steps that were manually performed in this work, e.g., the deployment of software to the automotive system, or the analysis of the resulting experiment data. 
As previously mentioned, additional follow-up studies would be the replication of this proof-of-concept using software and hardware closer to those adopted for consumer vehicles. 
That would require the software to abide at least partly to existing automotive regulations, and to run experiments on hardware facing resource constraints closer to what is currently present in real-world vehicles. 

\section*{Acknowledgment}
This work was supported by the project \emph{Highly Automated Freight Transports} (AutoFreight), funded by Vinnova FFI [2016-05413].

\bibliographystyle{splncs04}
\bibliography{library}

\end{document}